\documentclass[reprint,twocolumn,amsmath,amssymb,aps]{revtex4}
\usepackage{graphicx}
\usepackage{dcolumn}
\usepackage{bm}
\usepackage{amsmath}
\usepackage{amssymb}
\begin{document}

\title{The influence of oxygen and hydrogen adsorption on the magnetic
structure of the ultrathin iron film on the Ir(001) surface
}

\author{Franti\v{s}ek M\'aca$^{1}$}\email{maca@fzu.cz}
\author{Josef Kudrnovsk\'y$^{1}$}
\author{V\'aclav Drchal$^{1}$}
\author{Josef Redinger$^{2}$}
\affiliation{$^{1}$Institute of Physics ASCR, Na Slovance 2, CZ-182 21 Praha 8, Czech Republic\\
$^{2}$Department of Applied Physics, Vienna University of Technology,
Gusshausstr. 25/134, 1040 Vienna, Austria}

\date{\today}

\newpage

\begin{abstract}

We present a detailed ab initio study of the electronic structure and
magnetic order of an Fe monolayer on the Ir(001) surface
covered by adsorbed oxygen and hydrogen. The results are compared
to the clean Fe/Ir(001) system, where recent intensive studies
indicated a strong tendency towards an antiferromagnetic order and
complex magnetic structures. The adsorption of an
oxygen overlayer significantly increases interlayer distance between the
Fe layer and the Ir substrate, while the effect of hydrogen is much
weaker. We show that the adsorption of oxygen (and also
of hydrogen) leads to a p(2$\times $1) antiferromagnetic order of the Fe
moments, which is also
 supported by an investigation based
on a disordered local moment state.
Simulated
  scanning tunneling
images using the simple Tersoff-Hamann model hint that the
proposed  p(2$\times $1) antiferromagnetic order could be detected
even by non-magnetic tips.

\end{abstract}


\maketitle

\section{Introduction}

Magnetic overlayers, i.e., thin films of magnetic materials
on a nonmagnetic substrate, prepared by  molecular beam epitaxy,
represent a new type of material with potential technological
applications.
Among a number of such systems, recently the fcc-Fe/Ir(001) overlayer system has attracted
 both experimental and theoretical attention.
Thin overlayers were grown successfully with negligible Fe-Ir
intermixing.
A metastable unreconstructed monolayer (ML) of Fe on the Ir(001) surface
could be prepared \cite{feir_exp,verena}  and studied theoretically by ab initio
methods \cite{feir_th1,feir_th2}.
Both theoretical investigations predicted a complex magnetic ground state.
It should also be noted that complex chiral magnetic structures were
 also observed for a related bcc-Fe/W(001) system \cite{few_expth}.

Concerning adsorption, oxygen and hydrogen 
are the two most likely adsorbates
to exist on Fe/Ir(001) surfaces.
The investigation of their influence on the structural, electronic, and
magnetic ground state is thus an interesting and relevant problem.
We like to mention here a related recent experimental and theoretical study
of an oxygen covered bcc Fe(001) surface \cite{bcc_feo}, where
experiment confirms the formation of a well-defined monolayer where the oxygens
occupy the four-fold coordinated hollow positions and stabilize
the ferromagnetic (FM) state.
The situation is richer and even more interesting in the present
fcc Fe/Ir(001) case because oxygen adsorption can influence the Fe-Ir
distance and thus also the  magnetic ground state which sensitively
depends on it  \cite{feir_th1}.
In contrast, hydrogen which is likely to be present due to the
preparation of the  Fe/Ir(001) samples \cite{feir_exp}, adsorbs and hybridizes
differently from oxygen.

The ground state structure including possible layer relaxations
is determined by conventional ab initio tools (the supercell VASP method\cite{vasp1},
for more details see the next section).
The magnetic ground state is predicted on the basis of the
Heisenberg model with exchange interaction parameters, determined
from first-principles.
Two approaches are used: (i) in the first one,
two effective pair interactions are obtained utilizing the total energies
of the ferromagnetic (FM) and two antiferromagnetic (AFM) configurations (c(2$\times $2) and p(2$\times $1))
for the respective relaxed structural models calculated by VASP; and (ii)
the disordered-local moment (DLM) state is used as a reference state
for the extraction of exchange interactions as done in a previous paper
\cite{feir_th1} for the clean Fe/Ir(001) system.
The input electronic structure for the DLM state employs a realistic
semi-infinite geometry and the tight-binding linear muffin-tin orbital
(TB-LMTO) method in the Green function formulation.
The DLM state is simulated in the alloy analogy approximation
(for more details see Ref.~\onlinecite{feir_th1} and references
there).
We used the geometry as obtained by the atomic force minimization of VASP and chose
the  radii of the atomic spheres in order to minimize their overlap.
The above two approaches for the estimation of the exchange interactions are
complementary to each other, the former one is more accurate, but
since it employs only three magnetic configurations,
its predictions are limited to the simplest magnetic structures and more
complex configurations may thus be missed.
We demonstrate this for a monolayer of Fe on the Ir(001) surface.
The latter approach, based on the DLM method, does not assume any
specific magnetic structure and takes into account a large number of
exchange integrals (which in two-dimensional case decay more slowly
with distance than in the bulk case\cite{feco_cu}).
The DLM approach using the TB-LMTO method is, however, less accurate,
in particular for the case of small interlayer adsorbate-iron distances.
Finally, STM images simulating the constant current measuring mode of STM were
calculated using the Tersoff-Hamann  approximation\cite{TH85}, where tip-sample
interactions are neglected and thus the tip is supposed to follow contours of constant
density of states contained in an energy interval given by $E_{\rm F}$ and the applied
bias voltage $V_{\rm Bias}$.

\section{Computational details}

First principles density functional theory calculations were performed
using the Vienna ab-initio simulation package VASP \cite {vasp1},
using the projector augmented wave scheme \cite {vasp2}
and local density approximation (LDA) as given by Perdew-Zunger
(Ceperly-Alder) \cite {pzca}.
Repeated asymmetric slabs with seven layers of Ir and a single Fe
monolayer on one side and also symmetric slabs with eleven substrate
layers and adsorbate layers on both sides were used,  separated
by  at least 19~\AA ~vacuum.
Three top layer distances have been relaxed, the remaining
interlayer distances were fixed to the bulk experimental value for the Ir crystal
(1.92 \AA ). The calculated relaxations  turned out to be
essentially the same for both setups.
We have tested the FM-, c(2$\times $2)-AFM and p(2$\times $1)-AFM
arrangements, all performed with four Fe/Ir atoms in the two-dimensional supercell to guarantee a
reliable comparison of total energies.
Technically, we have used a Brillouin zone sampling with 100-200
special k-points in the irreducible two-dimensional wedge.
The difference between input and output charge densities in the final
iteration was better than 0.1 me bohr$^{-3}$.
The total force on single atoms was in every case smaller than
2.5~meV/\AA .

The input electronic structure for the DLM state was determined
for a realistic semi-infinite geometry in the framework of the
TB-LMTO method and the Green function formulation.
The same LDA functional as in VASP calculations was used \cite{pzca}
together with layer relaxations as obtained by VASP  and
atomic sphere radii chosen to minimize their overlap.
The vacuum above the overlayer was simulated by empty spheres (ES).
Electronic relaxations were allowed in three empty spheres layers
adjoining the oxygen overlayer,
in the oxygen and iron layers, and in five adjoining Ir substrate
layers.
This finite system was sandwiched selfconsistently between the frozen
semi-infinite fcc Ir(001) bulk and the ES vacuum-space including the
dipole surface barrier.
For more details see Ref.~\onlinecite{feir_th1} and references
there.

The effective exchange integrals $J_1$ and $J_2$ were estimated in
the {\em total energy model} from total energies calculated for three
different magnetic configurations as follows
\begin{eqnarray}
 J_1=\frac{1}{8}[E^{\rm tot}_{{\rm c}(2\times 2)}({\rm AFM})-
E^{\rm tot}_{{\rm c}(2\times 2)}({\rm FM})] \, , \nonumber \\
 J_2=\frac{1}{8}[E^{\rm tot}_{{\rm p}(2\times 1)}({\rm AFM})-
E^{\rm tot}_{{\rm p}(2\times 1)}({\rm FM})-4J_1] \, ,
\label{eqJen}
\end{eqnarray}
where corresponding energies were obtained for the calculated equilibrium
atomic structure and indicated magnetic phase.
We note that on the square lattice the most probable magnetic configurations are employed.

In the {\em DLM model} the exchange integrals $J^{\rm Fe,Fe}_{i,j}$ between
sites $i,j$ in the magnetic overlayer are expressed as follows (the generalized
Liechtenstein formula) (see e.g. \cite{eirev})
\begin{equation}
J^{\rm Fe,Fe}_{i,j}  =
\frac{1}{4 \pi} \, {\rm Im}
\int_{C} \, {\rm tr}_L
\left[ {\Delta^{\rm Fe}_{i}(z)} \,
g^{\uparrow}_{i,j}(z) \,
{\Delta^{\rm Fe}_{j}(z)} \,
g^{\downarrow}_{j,i}(z)
\right] \, {\rm d} z \, .
\label{eqJl}
\end{equation}
Here, the trace extends over $s-,p-$, and $d-$ basis set, the quantities
$\Delta^{\rm Fe}_{i}$ are proportional to the calculated exchange
splittings, and the Green function $g^{\sigma}_{i,j}$ describes the
propagation of electrons of a given spin ($\sigma=\uparrow,\downarrow$)
between sites $i,j$, both in the magnetic layer and via the Ir substrate.

Once the exchange interactions are known, we construct the corresponding
two-dimensional (2D) Heisenberg model in which only interactions between
iron moments are included explicitly and the interactions with iridium
atoms and with the adsorbate are included indirectly via selfconsistent
electronic structure calculations.
In the next step we determine its lattice Fourier transform J({\bf q})
using calculated exchange interactions up to 90 nearest-neighbors.
The wave vector ${\bf q}_0$ in the 2D Brillouin zone at which it acquires
its maximum determines the theoretically predicted magnetic ground state \cite{feir_th1}.

The STM simulations using the Tersoff-Hamman approach\cite{TH85} were done
with VASP using the optimum geometry as determined before. A fine k-point mesh
of 20$\times $40$\times $1 on an $\Gamma$-centered grid was used to sample the 2D Brillouin zone
and the energy cutoff was increased up to 600~eV
to ensure a smooth numerical representation even for the small values of the charge density
in the vacuum region. Energy slices according to the applied bias voltage, i.e. states between
$E_{\rm F}$ and $V_{\rm Bias}$, were considered.
A typical charge density contour of $10^{-6}$ puts a presumed tip at a distance of \mbox{$\sim$~3-4~\AA}~above the
terminal O atoms, a value not
to far from experimental distances. In order to simulate experimental conditions using
non-magnetic tips, the images are obtained from spin averaged densities.

\section{Results and discussion}

\subsection{Atomic structure}

In the following we present the results for possible ground state adsorbate structures.
All calculations here were done using the VASP-LDA method and the
the experimental bulk lattice constant for fcc Ir to facilitate a direct comparison
with our previous study \cite{feir_th1}.
This choice of exchange-correlation potential may lead to some underestimation of the
absolute values of interlayer distances for the adsorbates since
the experimental value (3.84 \AA ) is slightly larger than the theoretical one  (3.82 \AA ).
On the other hand, the trends like, e.g., an increase/decrease of
interlayer distances due to the adsorbate are predicted correctly.
In Table \ref{tab_geo} we summarize layer relaxations for various investigated systems.
Only the results for configurations minimizing the total energy are presented.

\begin{table}[ht]
\caption {\label{tab_geo}Calculated interlayer distances $d_{ij}$ between top
three sample layers for ground state magnetic configurations of Fe/Ir(001),
O on Fe/Ir(001), and H on Fe/Ir(001) in the AFM states. $b_i$ represents
the buckling of the atoms in the layer $i$.
Oxygen is adsorbed in a hollow site and hydrogen on favorable bridge position.
LEED experimental values are taken from Refs. \onlinecite{feir_exp, verena}.}
\begin{ruledtabular}
\begin{tabular}{l|l|l|l|l|l}
~&$d_{12} [\rm \AA ]$&$d_{23} [{\rm \AA }]$&$d_{34} [\rm \AA ]$&$d_{45} [\rm \AA ]$&$b_3 [\rm \AA ]$\\
\hline
c(2$\times $2)Fe&1.55&1.97&1.88&1.92&0.00\\
1 ML O on p(2$\times $1)Fe&0.59&1.98&1.86&1.90&0.00\\
0.5 ML O on p(1$\times $2)Fe&0.72&1.74&1.91&1.92&0.17\\
0.5 ML H on p(2$\times $1)Fe&1.12&1.59&1.92&1.92&0.17\\
\hline
Fe - exp.&1.69&1.96&1.91&1.92&0.00\\
H on Fe - exp.&-&1.72&1.94&1.91&0.00\\
\end{tabular}
\end{ruledtabular}
\end{table}

\begin{figure}[b]
\center \includegraphics[width=\columnwidth]{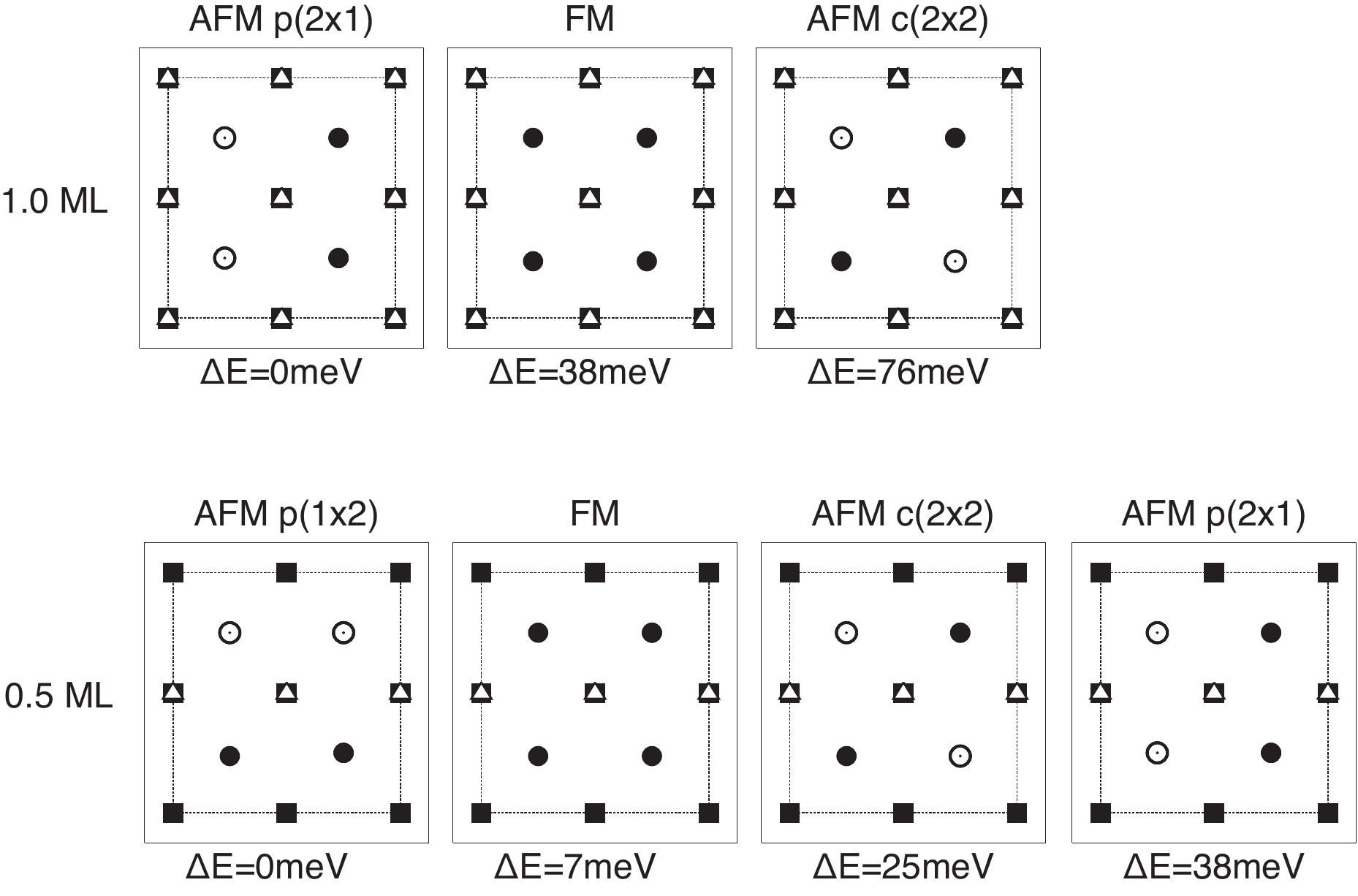}
\caption {Schematic structure of 1 ML  (top row) and 0.5 ML oxygen (bottom  row)  on a magnetic Fe/Ir(001) surface.
Squares mark the topmost Ir substrate atoms, circles Fe atoms (full/empty denote up/down spin) and white triangles the oxygen atoms.
The energy difference per Fe atom with respect to the lowest energy solution is indicated beneath each configuration.
}
\label{fig_enO}
\end{figure}

It is well known that oxygen atoms favor the occupation of the fourfold hollow Fe sites  (see, e.g., Ref. \onlinecite{bcc_feo}
and references therein). This site preference is also reproduced
by of our VASP calculations, which put the O atoms directly above the substrate Ir atoms.
Fig. \ref{fig_enO} shows that for 1 ML oxygen coverage a p(2$\times $1) AFM magnetic
order of Fe magnetic moments is preferred
(E$_{\rm AFM}^{{\rm c}(2\times 2)} >$ E$_{\rm FM} > $ E$_{\rm AFM}^{{\rm p}(2\times 1)}$,
$\Delta $E$_{\rm tot}$ = E$_{\rm FM}$ - E$_{\rm AFM}^{{\rm p}(2\times 1)}$ = 38 meV/Fe-atom).
For a 0.5 ML oxygen coverage diverse geometrical arrangements are possible.

We considered c(2$\times $2)O and p(2$\times $1)O adlayers and compared minimum
total energies (E$_{\rm min}$) for the FM and three different AFM configurations:
c(2$\times $2) AFM,  p(2$\times $1) AFM and p(1$\times $2)
AFM magnetic ordering of Fe moments.
Oxygen atoms prefer a p(2$\times $1) type overlayer structure
($\Delta $E$_{\rm tot}$ = E$_{\rm min}^{{\rm c}(2\times 2){\rm O}}$ - E$_{\rm min}^{{\rm p}(2\times 1){\rm O}}$ = 65 meV/O-atom),
the lowest total energy has been found for the p(1$\times $2) AFM magnetic ordering of Fe moments
(E$_{\rm AFM}^{{\rm p}(2\times 1)} >$ E$_{\rm AFM}^{{\rm c}(2\times 2)} >$ E$_{\rm FM} > $ E$_{\rm AFM}^{{\rm p}(1\times 2)}$,
$\Delta $E$_{\rm tot}$ = E$_{\rm FM}$ - E$_{\rm AFM}^{{\rm p}(1\times 2)}$ = 7 meV/Fe-atom).

\begin{figure}[h]
\center \includegraphics[width=\columnwidth]{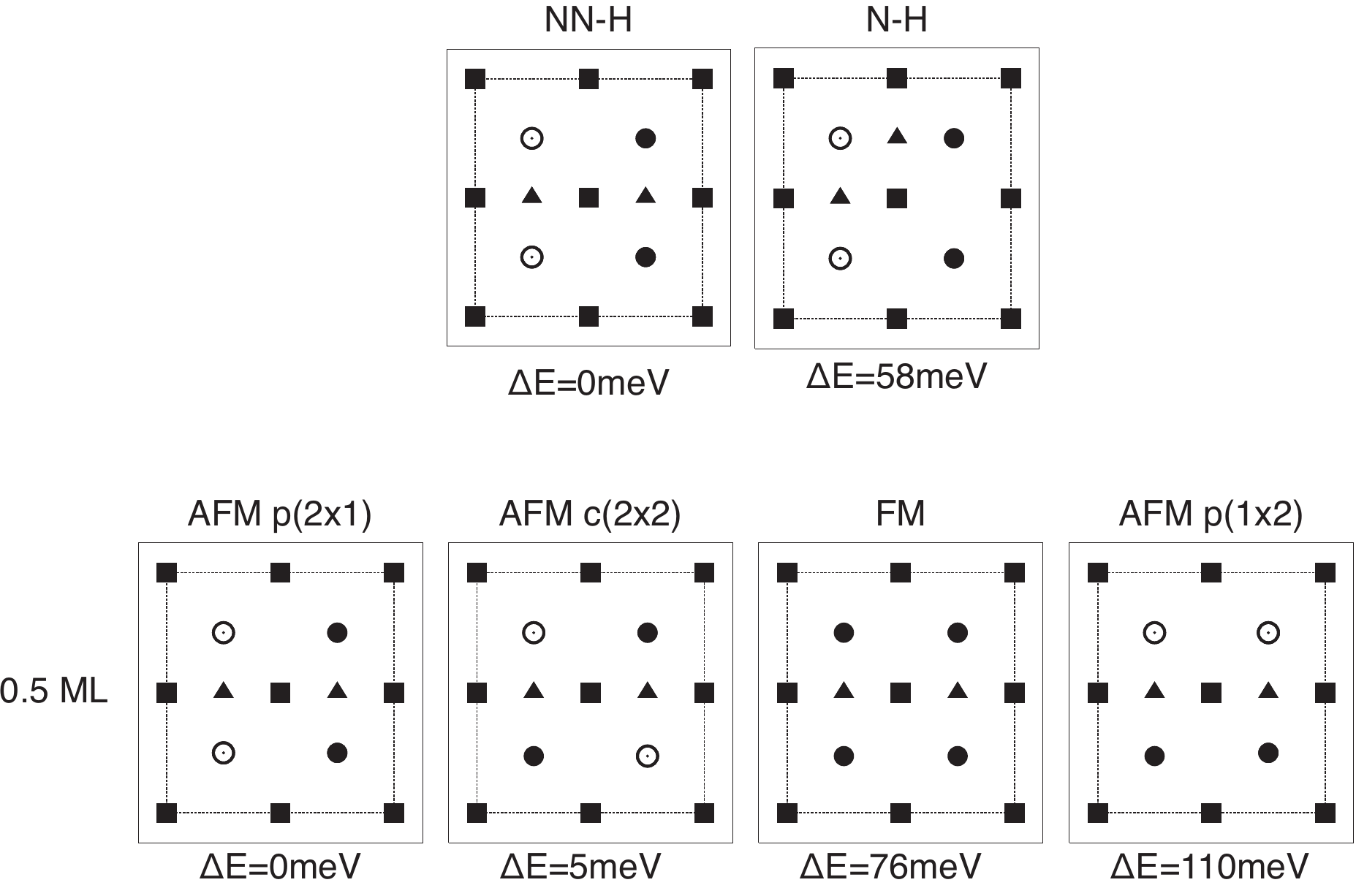}
\caption {
Schematic structure of 0.5 ML hydrogen occupying Fe-bridge sites on a magnetic Fe/Ir(001) surface.
Squares mark the topmost Ir substrate atoms, circles Fe atoms (full/empty denote up/down spin) and white triangles the hydrogen atoms.
Upper panel: Hydrogen atoms occupying either
next-neighboring (NN-H) Fe-bridge sites (left frame) or less stable neighboring sites (right frame) on a p(2$\times $1) antiferromagnetic Fe/Ir(001) surface.
Lower panel: Different magnetic configurations for hydrogen occupying next-neighboring Fe-bridge sites.
The energy differences $\Delta E$ with respect to the minimum energy solution is given per H atom  in the upper  and
per Fe atom in the lower panel.
}
\label{fig_enH}
\end{figure}

 In contrast to oxygen, hydrogen atoms prefer Fe-bridge sites on the Fe/Ir(001) surface as shown schematically in Fig. \ref{fig_enH}.

Similar to oxygen adsorbates, different geometrical arrangements are possible for a 0.5 ML coverage.
In our model calculations
we considered two H adlayer geometries, H atoms occupying neighboring (N-H)
or next-neighboring Fe-bridge (NN-H) sites.
Again the energies for FM and three AFM ordered states are compared, i.e. for a c(2$\times $2) AFM,
a p(2$\times $1) AFM, and a p(1$\times $2) AFM magnetic ordering of Fe moments.
Our calculations show that H atoms prefer to occupy next-neighboring (NN-H)  Fe-bridge
sites ($\Delta E_{\rm tot}$ = $E_{\rm min}^{\rm N-H}$ - $E_{\rm min}^{\rm NN-H}$ = 58 meV/H-atom) and
the ground state is obtained for a p(2$\times $1) AFM magnetic ordering
of Fe moments (H atoms between parallel Fe-moments).  Fig.  \ref{fig_enH} shows the magnetic ordering with
E$_{\rm AFM}^{{\rm p}(1\times 2)} >$  E$_{\rm FM} > $ E$_{\rm AFM}^{{\rm c}(2\times 2)} >$ E$_{\rm AFM}^{{\rm p}(2\times 1)}$,
and $\Delta $E$_{\rm tot}$ = E$_{\rm AFM}^{{\rm c}(2\times 2)}$ - E$_{\rm AFM}^{{\rm p}(2\times 1)}$ = 5 meV/Fe-atom).
The optimized geometry further showed that an unsymmetrical (N-H) occupation of bridge sites
on the Fe/Ir(001) surface induces considerable buckling in the subsurface Ir layer
($b_3$ = 0.17~\AA ).

We suppose that under real conditions the hydrogen adsorption starts for lower coverage
($\Theta \approx 0$) with a random occupation of bridge sites.
By increasing the H coverage  ($\Theta \rightarrow 1$) the ordering of adsorbed H atoms
starts to support the p$(2 \times 1)$ magnetic structure.
Nevertheless, the formation of domains can be expected rather than a long-range order.

\subsection{Electronic structure}

Here we wish to discuss the changes induced by hydrogen and oxygen
 adsorption on the  Fe/Ir(001) surface as reflected in the local density
of states (LDOS).
We only show results for the structural and magnetic configurations
with the lowest energy as determined in the previous subsection.
The changes induced by the H- and O-adsorption is best discussed
by comparing with the clean reference Fe/Ir(001) system for which in
in Refs.~\onlinecite{feir_th1,feir_th2}
a complex magnetic ground state  (chiral magnetic structure) was predicted.

\begin{figure}[h]
\center \includegraphics[width=0.75\columnwidth]{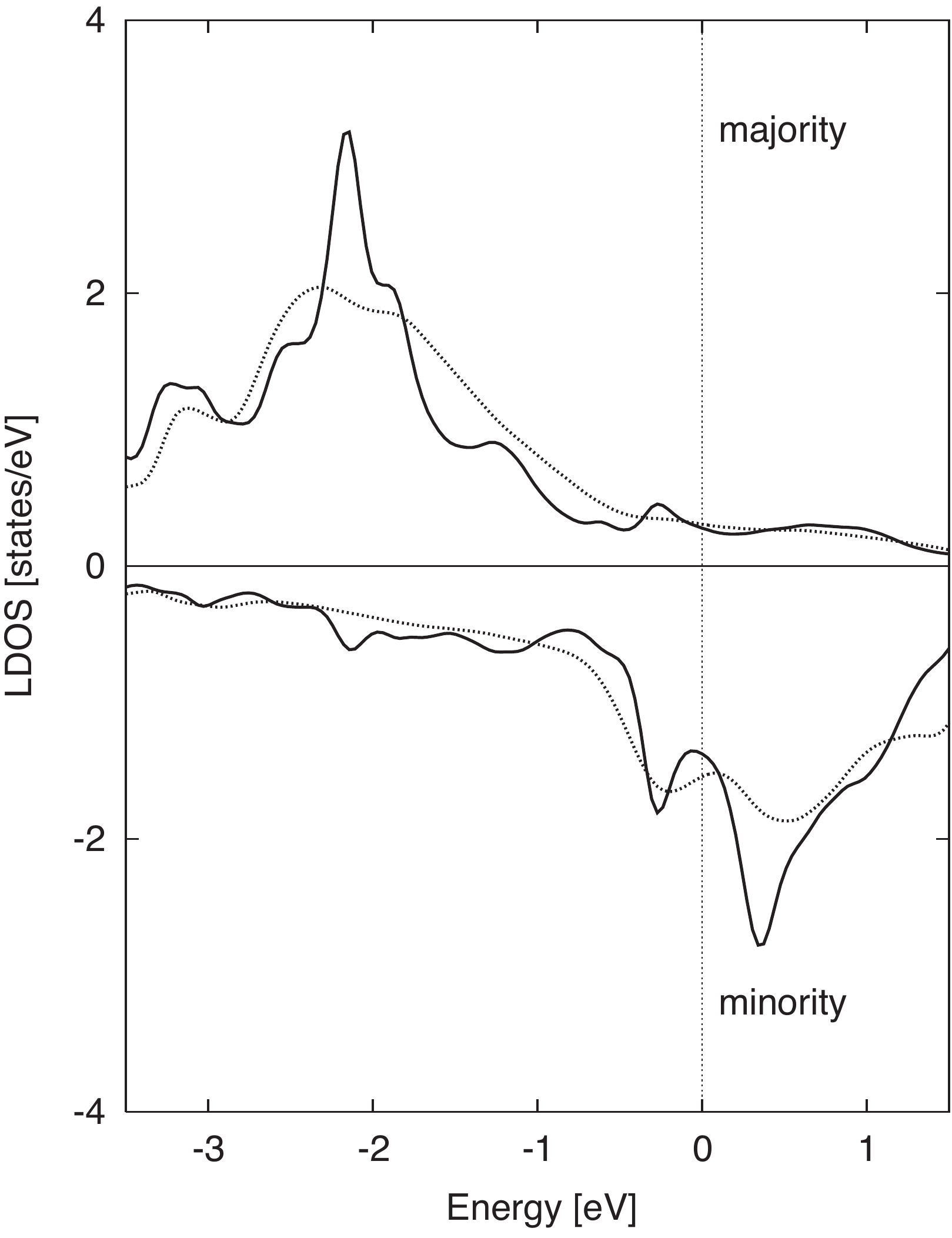}
\caption {Spin resolved local density of states (LDOS) of a Fe atom in of Fe/Ir(001):
(i) c(2$\times $2) AFM (full lines), and (ii) DLM
solution (dotted lines) for the experimental geometry and $E_{\rm F}$ set to zero.
}
\label{fig_com}
\end{figure}

In Fig.~\ref{fig_com} we compare the LDOS's for a c($2\times2)$ AFM
 ordered Fe on Ir(001) with the Fe/Ir(001) system in the DLM state,
which can be considered as a 'disordered' AFM state.
In order to exclude the effects of different interlayer distances, the same
geometry was used in both cases.
We  find a good overall agreement between both LDOSs (main peaks, the widths),
which means that if we find relevant changes between the reference
Fe/Ir(001) system and its counterpart with adsorbed  O- or H-atoms, these
changes reflect mostly changes induced by adsorbate-Fe hybridization
\cite{bcc_feo} rather than small differences between similar
magnetic configurations.

\begin{figure}[h]
\center \includegraphics[width=0.75\columnwidth]{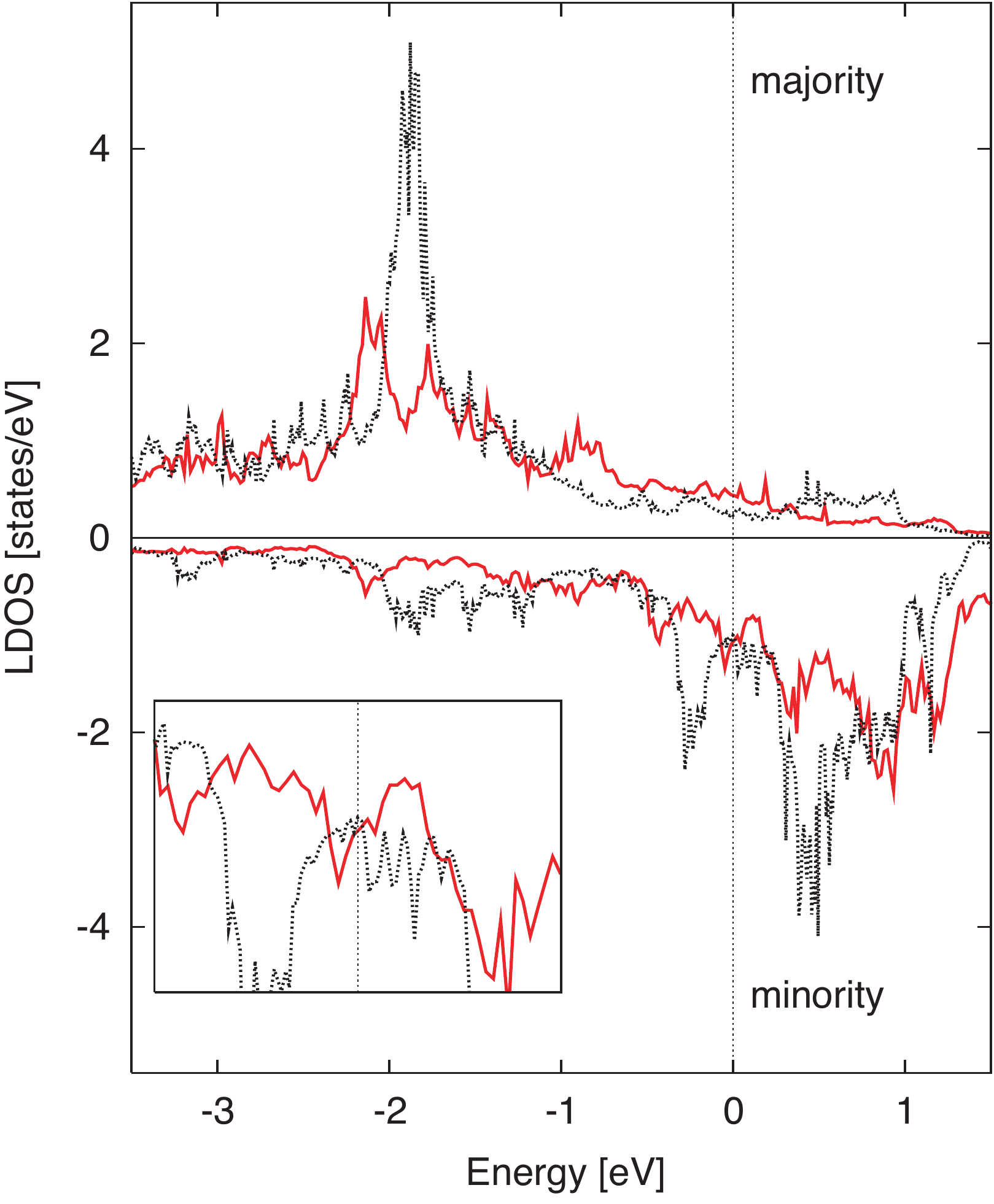}
\caption {(Color online) Spin resolved local density of states (LDOS) of an Fe atom
of antiferromagnetic p(2$\times $1) Fe/Ir(001) covered
by an oxygen monolayer (full lines)  compared to the clean
antiferromagnetic c(2$\times $2) Fe/Ir(001) system (dotted lines).
The inset shows the vicinity of the Fermi level ($E_{\rm F}$=0) for minority
states.
}
\label{fig_dosO}
\end{figure}

\begin{figure}[h]
\center \includegraphics[width=0.75\columnwidth]{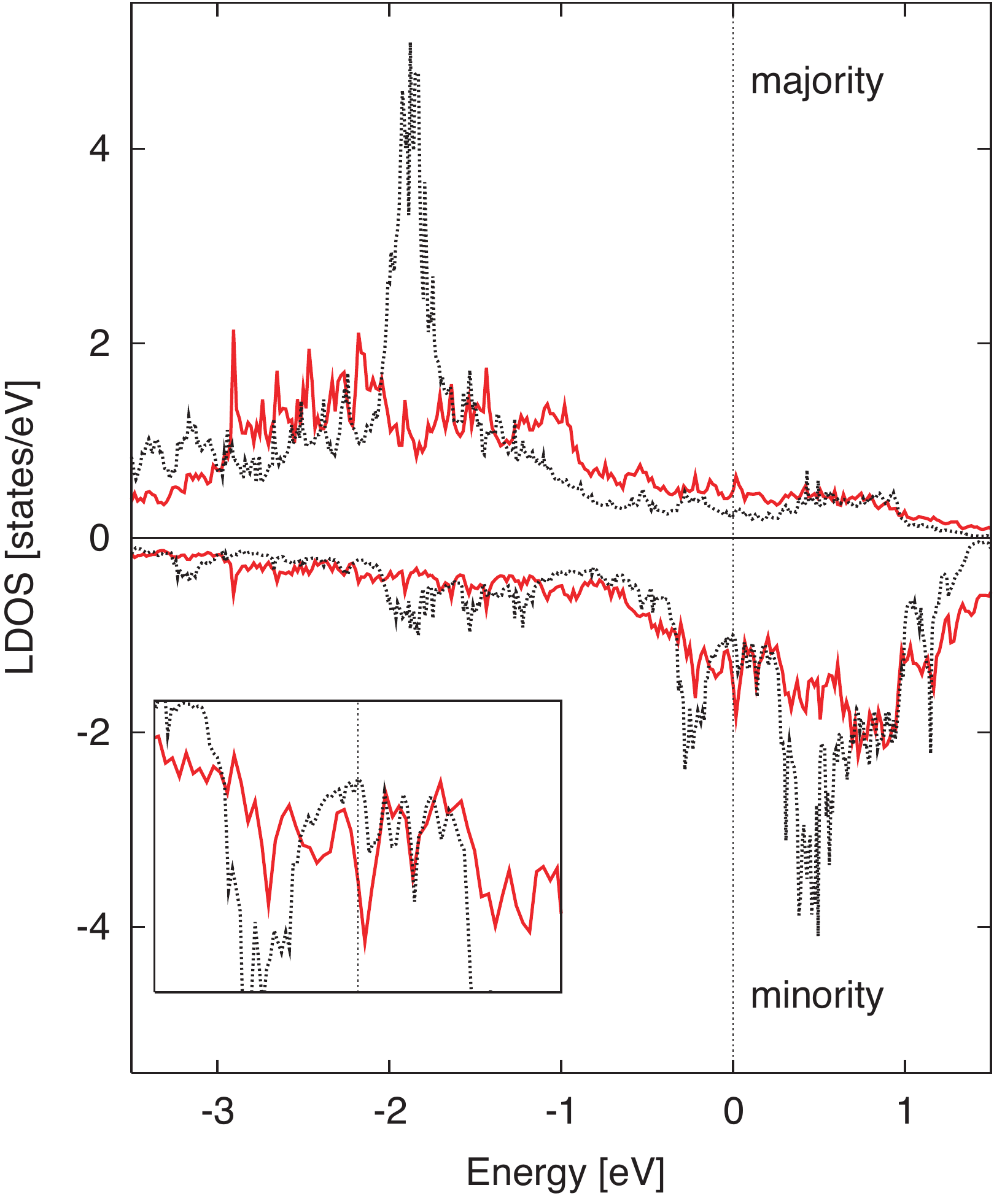}
\caption {(Color online) The same as in Fig.~\ref{fig_dosO} but for a
 p(1$\times $2) ordered  half-monolayer oxygen coverage (full lines).
}
\label{fig_dos0.5O}
\end{figure}

The effect of the oxygen adsorption on Fe/Ir(001) is shown in
Fig.~\ref{fig_dosO} for 1 ML O and it is compared with
the reference c(2$\times $2) Fe on the Ir(001) surface (no oxygen).
In both cases fully relaxed geometries were used.
The most important effect shown in Fig. \ref{fig_dosO} is a strong change of
the LDOS features due to the oxygen-iron hybridization.
The effect is stronger for majority states as compared to minority states
similar to  oxygen-covered bcc-Fe(001) \cite{bcc_feo}.
In Fig.~\ref{fig_dos0.5O} we show the LDOS for an ordered half-monolayer
of oxygen.

While the overall shape of the LDOS is similar for the two oxygen coverages,
some changes can be traced down in the vicinity of the Fermi level
(see insets in Figs.~\ref{fig_dosO},~\ref{fig_dos0.5O} showing in detail
the density for minority states).
The  changes in the LDOS due to the oxygen-iron wave function
hybridization as described above are a precursor of differences in
their magnetic stability (see Fig.~\ref{fig_jqO} below).
\begin{figure}[h]
\center \includegraphics[width=0.75\columnwidth]{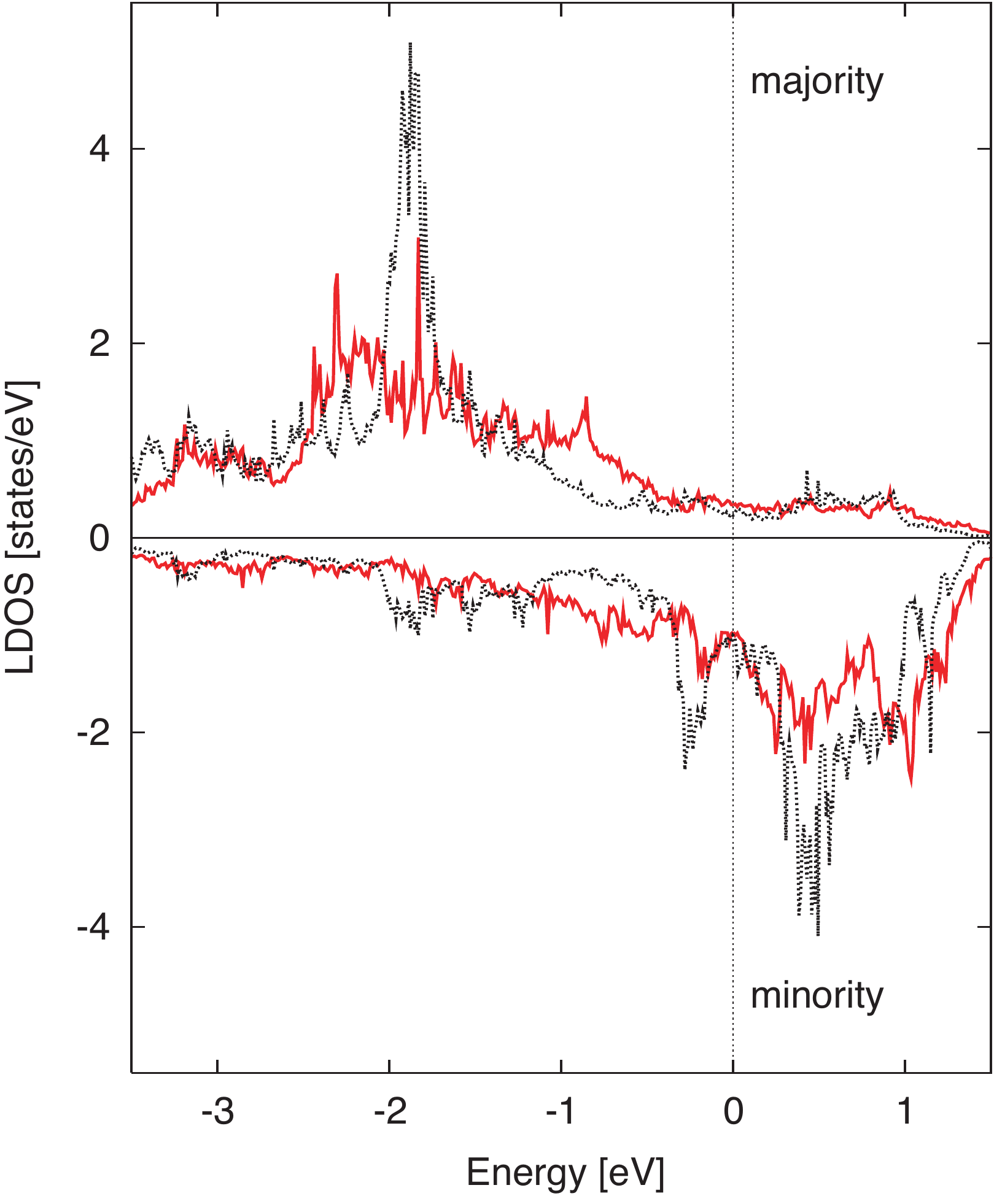}
\caption {
(Color online) Spin resolved local density of states (LDOS) of an Fe atom
of antiferromagnetic p(2$\times $1) Fe/Ir(001) covered
by half a monolayer of hydrogen (full line) compared to the clean
antiferromagnetic c(2$\times $2) Fe/Ir(001) system (dotted lines)
and $E_{\rm F}$ set to zero.
}
\label{fig_dosH}
\end{figure}

The effect of hydrogen adsorption on the LDOS is similar to oxygen as
shown for the adsorption  of 0.5 ML H in Fig.~\ref{fig_dosH}.
We again see mostly a broadening of LDOS-peaks due to the Fe-H interaction.

\subsection{Magnetic phase stability}

\begin{figure}
\center \includegraphics[width=0.75\columnwidth]{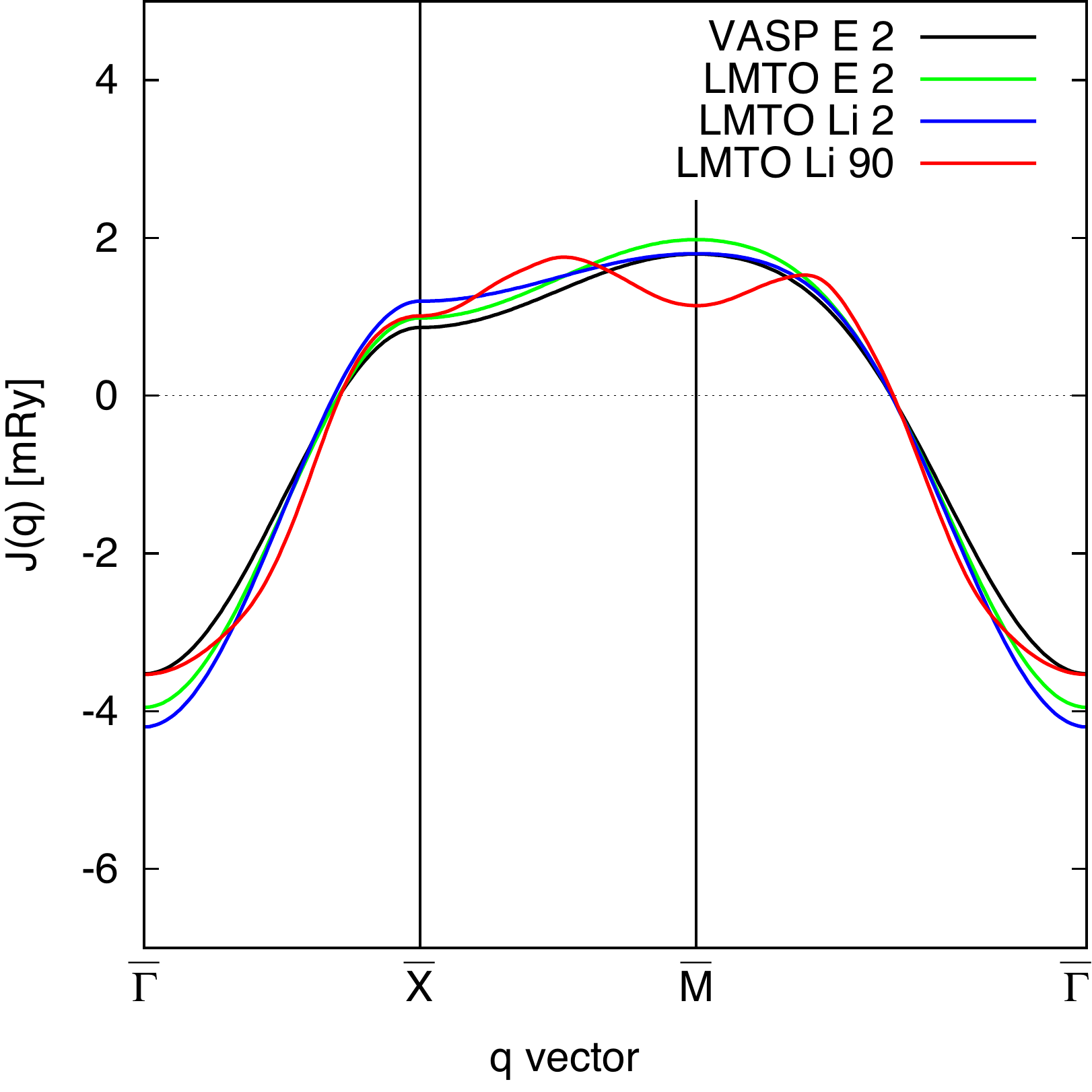}
\caption {(Color online) Lattice Fourier transform of the real-space exchange
interactions $J^{\rm Fe,Fe}_{ij}$, $J$({\bf q$_{\|}$}), for the
clean reference Fe/Ir(001) system in the relaxed minimum energy geometry
obtained from
total energy  
 and DLM approaches.   Total energy approach: using VASP (VASP E 2) and
TB-LMTO (LMTO E 2). DLM approach using TB-LMTO and just 2 exchange integrals
(LMTO Li 2) or TB-LMTO with 90 exchange integrals
(LMTO Li 90).
}
\label{fig_jqref}
\end{figure}

We study the magnetic phase using the lattice Fourier transform of the
real-space exchange integrals estimated using the total energy and DLM models, as
shown in Fig.~\ref{fig_jqref}.
We start with the reference case of the Fe/Ir(001) system studied previously
in the framework of the DLM model \cite{feir_th1}, but compare it here
with its total energy model counterparts (see Eq.~(\ref{eqJen}). The corresponding
total energies have been obtained both from VASP and TB-LMTO calculations.
The DLM approach is shown here for two cases with vastly different numbers
of exchange integrals included in the lattice Fourier transform, namely
for 90 integrals and for just the two leading terms (two shells of neighbors).
It should be noted that the DLM exchange integrals correspond to bare
ones while those obtained from total energies are effective ones which
in some sense combine all of them into two terms.

We see an overall good agreement between both approaches, nevertheless  
in the full DLM approach (90 shells) the ground state moves from the antiferromagnetic c(2$\times $2)
state (the ${\bar {M}}$-ordering vector) to a more complex magnetic
state with an ordering vector lying on the ${\bar {X}}$ - ${\bar {M}}$ line.
We also note that both  the p(2$\times $1) AFM state (the ${\bar {X}}$-ordering
vector) and c(2$\times $2) AFM state are energetically close while the ferromagnetic state (the
${\bar {\Gamma}}$ ordering vector) is energetically higher (due to
the adopted convention for the Heisenberg Hamiltonian \cite{feir_th1}
the lowest energy corresponds the maximum of the $J$(${\bf q_{\|}})$ curve).

\subsubsection{O- and H-adsorption: Total energy model}

\begin{figure}[h]
\center \includegraphics[width=0.75\columnwidth]{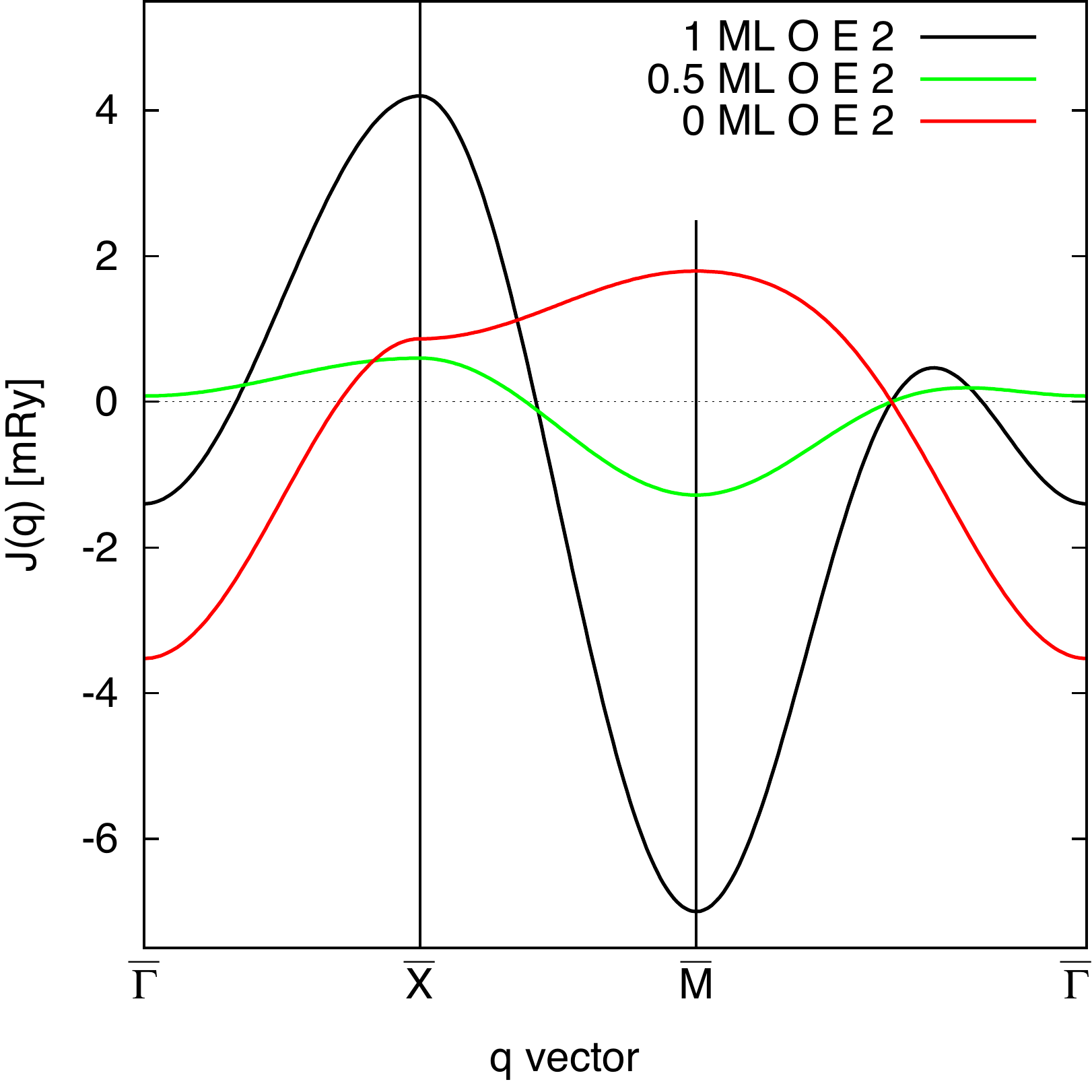}
\caption {(Color online) Lattice Fourier transform of real-space exchange
interactions $J^{\rm Fe,Fe}_{ij}$, $J$({\bf q$_{\|}$}), for
clean antiferromagnetic c(2$\times $2) Fe/Ir(001) (0 ML O E 2) and covered
by half a monolayer of oxygen (0.5 ML O E 2) and an oxygen monolayer (1 ML O E 2).
The oxygen atoms occupy fourfold Fe hollow sites.
}
\label{fig_jqO}
\end{figure}

The lattice Fourier transform for the oxygen adsorption is shown in
Fig.~\ref{fig_jqO} and compared with the reference oxygen-free case.

We see that  1ML  oxygen induces a p(2$\times $1) antiferromagnetic
 state (the ${\bar {X}}$-ordering vector) in contrast to the c(2$\times $2)
 antiferromagnetic state (the ${\bar {M}}$-ordering
vector) originally considered for the clean surface.
The high stability of the p(2$\times $1) state is strongly reduced for
a half-monolayer oxygen coverage.
Interestingly the ferromagnetic state is energetically
rather close to the ground state, which can be ascribed to the interplay between Fe-O hybridization
which promotes p(2$\times $1) state and the increase of the Fe-Ir distance upon O adsorption.
The ferromagnetic state is the ground state
for large Fe-Ir interlayer distances in the Fe/Ir(001) system \cite{feir_th1}.

\begin{figure}[h!]
\center \includegraphics[width=0.75\columnwidth]{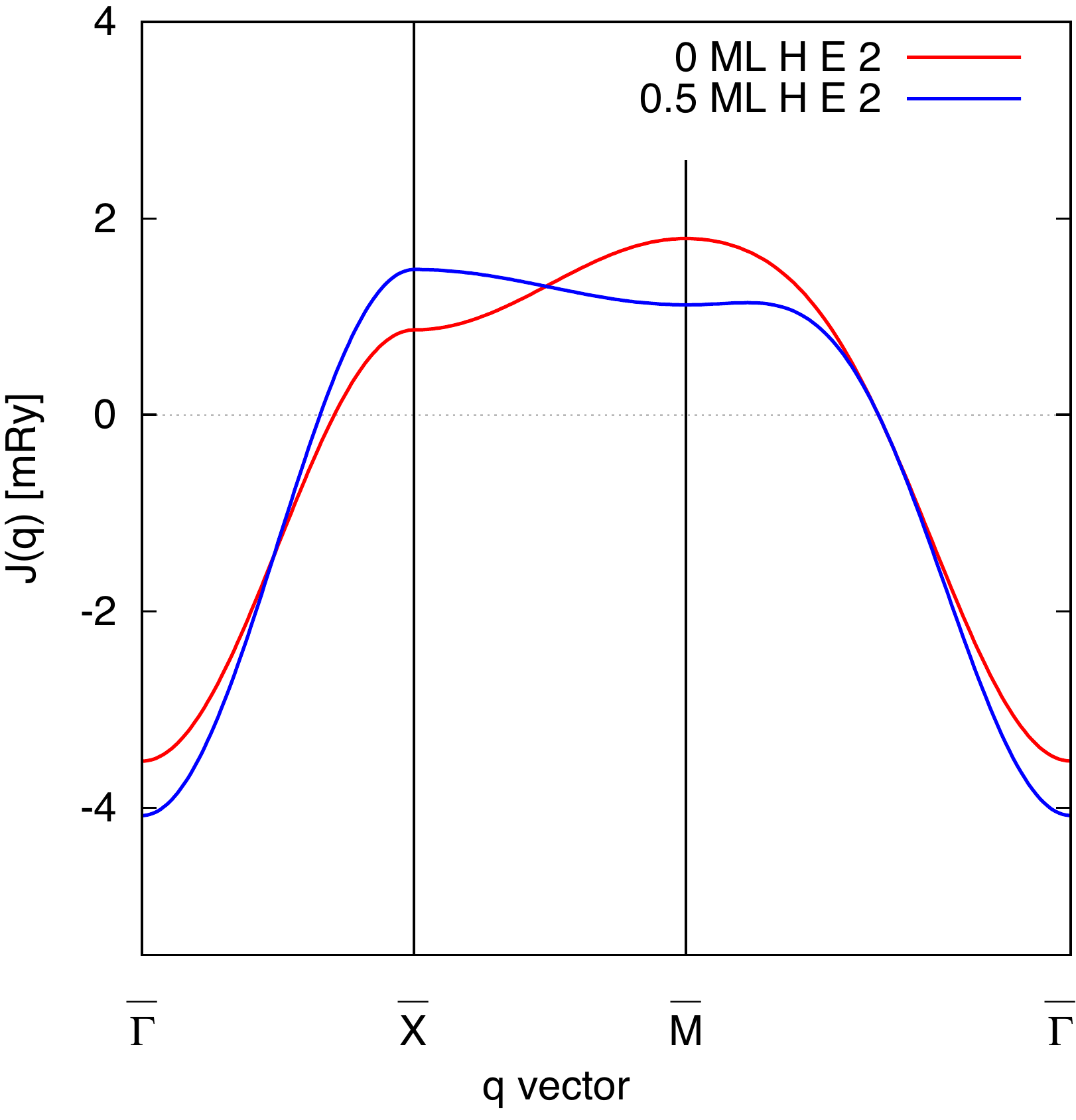}
\caption {(Color online) Lattice Fourier transform of real-space exchange
interactions $J^{\rm Fe,Fe}_{ij}$, $J$({\bf q$_{\|}$}), for the
Fe/Ir(001) system covered by half a monolayer of hydrogen (0.5 ML H E 2).
 The magnetic ground state is obtained for an
antiferromagnetic p(2$\times $1) ordering  of Fe the moments.
The hydrogen-free clean antiferromagnetic c(2$\times $2) Fe/Ir(001) system (0 ML H E 2)
is shown for comparison.
}
\label{fig_jqH}
\end{figure}

The energetic ordering can be understood, at least qualitatively, by applying the
Kanamori-Goodenough-Anderson \cite{kanamori, goodenough, anderson} rules in a simple way.
These rules predict that Fe-O-Fe bond angles of 90$^\circ$ favor
a ferromagnetic (FM) coupling of  the Fe atoms while Fe-O-Fe bond angles of 180$^\circ$ favor an
antiferromagnetic (AFM) one. We apply now these rules for the 1~ML O overlayer as depicted in the top
panel of Fig. \ref{fig_enO}. The preferred magnetic coupling is a result of magnetic interactions between neighboring Fe atoms (Fe-O-Fe bond angles 85$^\circ$)
and next-neighboring Fe atoms (Fe-O-Fe bond angles 146$^\circ$). For the
present purpose, the values of 85$^\circ$ and 146$^\circ$
for the calculated minimum energy geometry are sufficiently close to the ideal values
of 90$^\circ$ and 180$^\circ$.  Summing up neighboring and
next-neighboring Fe couplings one easily finds that for the most stable AFM p(2$\times $1) state
one half of the 85$^\circ$ couplings are favorable while the other half is unfavorable,
whereas both 146$^\circ$ couplings are favorable.
In case of the FM configuration the favorable 85$^\circ$ couplings are offset by unfavorable 146$^\circ$ FM
couplings. For AFM c(2$\times $2) all couplings are unfavorable, indicating a predominance of Fe-O-Fe
induced couplings over the substrate related c(2$\times $2) AFM coupling of the Fe atoms, which is also offset
by the increased Fe/Ir distance upon O adsorption.
In summary one can safely state that oxygen adsorption on the Fe/Ir(001) surface strongly influences the magnetic
ordering in the Fe monolayer.
Turning now to the influence of hydrogen adsorbed at the preferred Fe-bridge sites
on the magnetic state, the results for the lattice Fourier transform are summarized
in Fig.~\ref{fig_jqH}.

A weak preference for the antiferromagnetic p(2$\times $1)
state (${\bar {X}}$-ordering vector) is seen, as expected from the total energy
calculations but the antiferromagnetic c(2$\times $2) state (${\bar {M}}$-ordering vector)
is energetically quite close.
These results suggest, that the effect of H-adsorption on the magnetic state
of Fe/Ir(001) is weaker as compared to the O-adsorption.
This is also accompanied by a weaker influence on the geometrical structure.

\subsubsection{O-adsorption: DLM model}

\begin{figure}
\center \includegraphics[width=0.75\columnwidth]{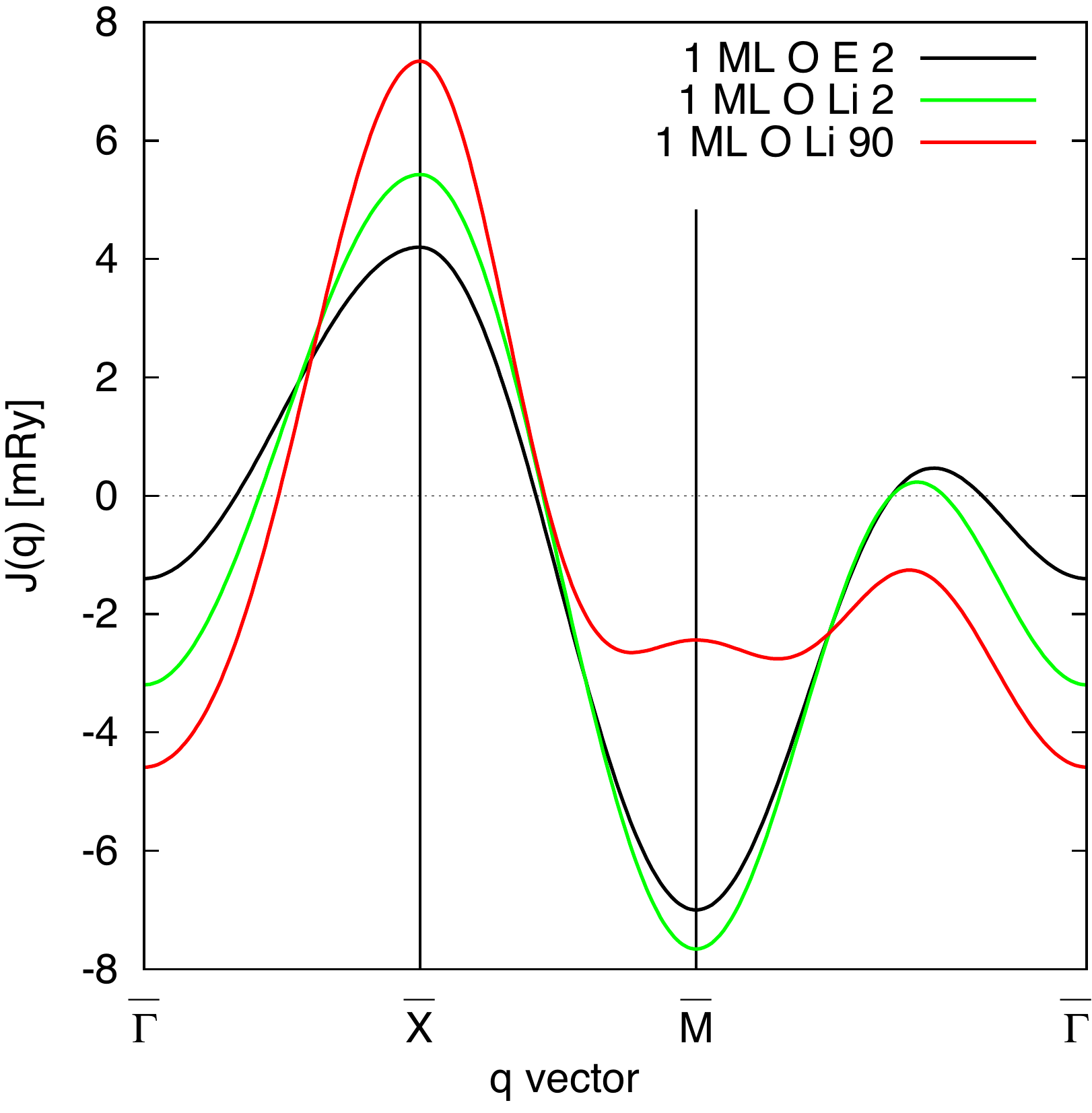}
\caption {(Color online) Lattice Fourier transform of the real-space exchange
interactions $J^{\rm Fe,Fe}_{ij}$, $J$({\bf q$_{\|}$}) for the
 Fe/Ir(001) covered by the 1 ML of oxygen. determined
in the framework of the DLM approach is compared to results from a total
 energy approach using VASP (1 ML O E 2). Compared are two different TB-LMTO based
  DLM approaches, one with just two
exchange integrals (LMTO Li 2) and the other with all 90 calculated exchange
integrals (LMTO Li 90).
}
\label{fig_jq_dlmO}
\end{figure}

In this subsection we present results for the magnetic stability
of the oxygen monolayer as determined from the DLM model.
We wish to point out that the TB-LMTO-DLM model employs the spherical
charge approximation which is less accurate as compared to the full
potential methods like VASP.
We have shown, however, that for layer relaxations up to 15\% of the
interlayer distance (the case of Fe/Ir(001) \cite{feir_th1}) the results
obtained using the total energy and TB-LMTO methods are in a good agreement
(see Fig.~\ref{fig_jqref} and discussion there).
However, the situation for an O-monolayer is, worse as the O-Fe interatomic distance
amounts to only 2.01~\AA ~(as compared to the host interatomic
distance of 2.72~\AA).
The problem in our TB-LMTO model consists in the division of the space into vacuum, oxygen,
and iron spheres which is not unique.
We have therefore used the same approach as in our previous work
\cite{feir_th1}.
A similar approach used recently for the oxygen adsorption on
bcc-Fe(001) has lead to a qualitative agreement with experiment.
In Fig.~\ref{fig_jq_dlmO} we present results for 1ML O on Fe/Ir(001) and compare
them with the results of the supercell VASP
approach as presented in Fig.~\ref{fig_jqO}.

Like in Fig. \ref{fig_jqref}, we tested the dependence of the
magnetic stability  on the
number of shells included in calculations of the lattice Fourier transform.
The most important conclusion, namely, that the oxygen monolayer
stabilizes the antiferromagnetic p(2$\times $1) ground state for
Fe/Ir(001), was unambiguously obtained in both approaches.
In addition, even quantitative agreement is satisfactory although
the differences for various number of shells included in the DLM
method are larger as compared to the oxygen free case (see
Fig.~\ref{fig_jqref}).
On the other hand, there is no indication for a more complex magnetic
state if the number of included exchange interactions is increased.

\subsection{Scanning tunneling microscopy}

\begin{figure}
\center \vspace{0.25cm} \includegraphics[width=\columnwidth]{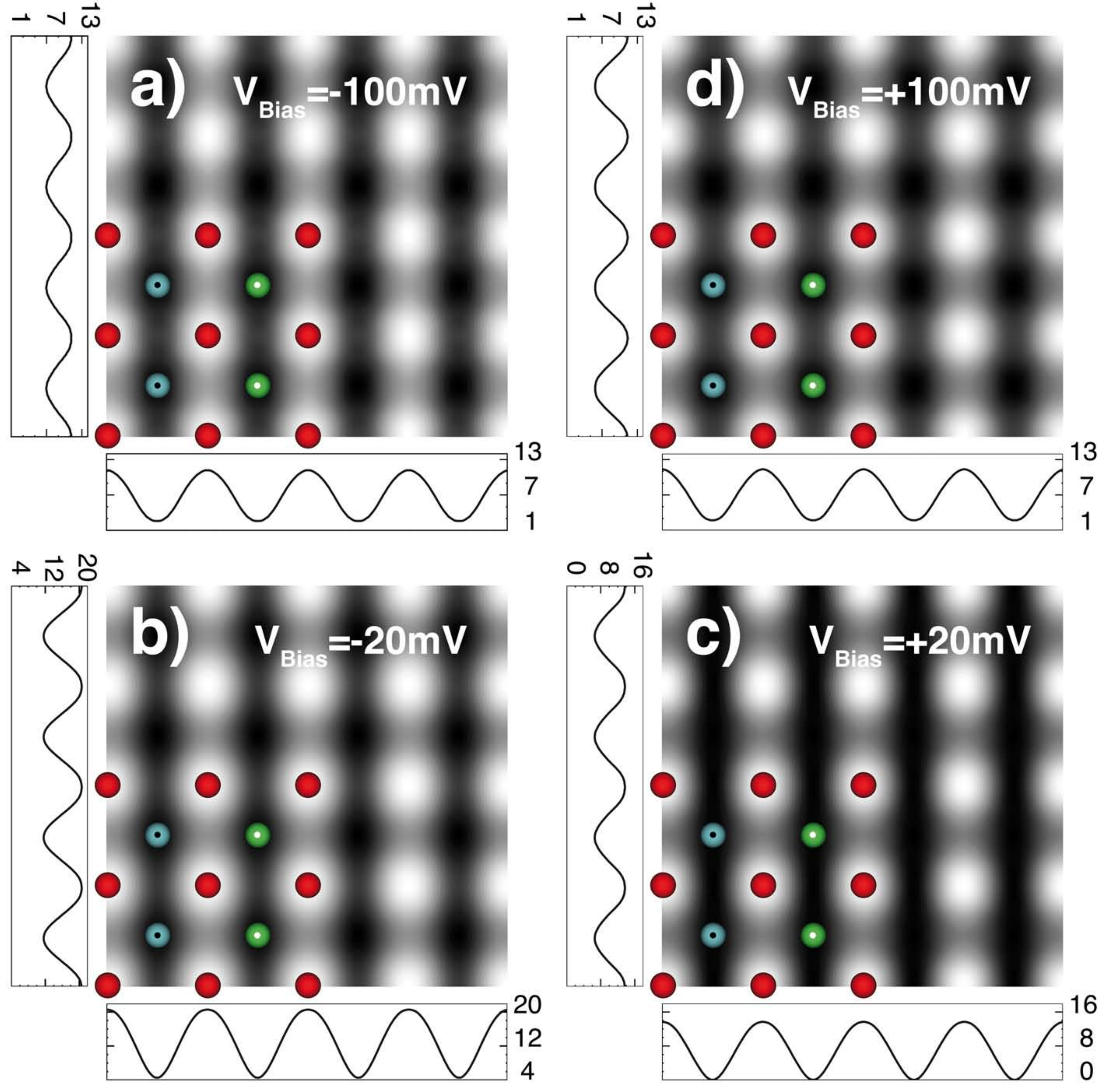}
\caption {(Color online) Simulated (Tersoff-Hamann model) constant current STM images of  1ML O on an antiferromagnetic p(2$\times $1)
Fe/Ir(001) surface for bias voltages $V_{\rm Bias}$ around $E_{\rm F}$ (a-d).
Red spheres denote the positions of the O atoms, blue and green spheres mark
 Fe atoms with up and down moments, respectively. Simulated line scans across the oxygen rows are drawn
 outside each image. A different corrugation for scans either parallel to FM or AFM
 ordered Fe atom rows is clearly visible, highlighting the p(2$\times $1) AFM ordering.
 The chosen charge density contour leads to tip-sample distances of $\sim$~3-4~\AA~(above the
O atoms).
}
\label{fig_stm}
\end{figure}

Surprisingly, the simulated constant current STM images shown in Fig.~\ref{fig_stm}
resolve the p(2$\times $1) AFM magnetic ordering of the Fe atoms even for a simulation
assuming  an  unpolarized tip in the Tersoff-Hamann model.
Although a spin polarized tip could directly deal with the different orientation
of the Fe moments, the AFM ordering of
the Fe atoms as imprinted on the O atoms by hybridization
may be also detected by a non-magnetic tip.
 For small bias
voltages, $\pm$~100~meV around E$_F$, the consequences of magnetic Fe-O hybridization
are clearly visible in the corrugation of the O atoms. Scanning above
oxygen rows either parallel to FM oriented Fe rows or AFM oriented Fe rows,
leads to corrugation differences of up
to 8 pm,  as shown in the simulated line scans in Fig.~\ref{fig_stm}. Resolving
such differences is certainly possible for experimental STM scans and thus should
yield a direct information on the predicted magnetic ordering.


\section{Conclusions}

We have investigated the effect of oxygen and hydrogen adsorption
on the structural and magnetic properties of Fe/Ir(001) system
from first-principles.
While the structural part was solved using an supercell VASP approach,
 we used two complementary approaches for the prediction
of the magnetic state:
A total energy model using supercell VASP calculations and the
TB-LMTO-DLM method working with the semi-infinite geometries.
The emphasis was put on the influence of the O and H adsorption on the
magnetic stability.
The following main conclusions can be drawn:
(i) Oxygen adsorbs on the Fe/Ir(001) surface at four-fold hollow Fe sites
and influences the atomic geometry (interlayer distances) of the system more
strongly than hydrogen adsorbing at Fe bridge positions. In particular, the adsorption of an
oxygen monolayer strongly increases the interlayer distance between Fe and top Ir layer;
(ii) The oxygen-iron hybridization mainly broadens density of states features
as compared to the clean surface. This change of the electronic
structure manifests itself in corresponding modifications of exchange
interactions and the STM current;
(iii) The magnetic stability is influenced by oxygen adsorption, and
we predict an antiferromagnetic p(2$\times $1) magnetic ground state
 with a ${\bar {X}}$-ordering vector as
 obtained by two complementary approaches, namely
 supercell VASP and DLM model;
(iv) The ${\bar {X}}$-ordering is weakened by decreasing oxygen coverage
and changes into a complex magnetic ground state for oxygen-free Fe/Ir(001);
(v) Hydrogen adsorption leads to a weak stabilization of
an antiferromagnetic p(2$\times $1) order; and
(vi) STM images for nonmagnetic tips reflect the p(2$\times $1) AFM ordering of the Fe moments not
directly, but rather by hybridization with the O atoms, best visible for small bias voltages.

\begin{acknowledgments}
F.M., J.K., V.D. acknowledge support of the Czech Science Foundation
(Projects IAA100100912 and P202/09/0775).
J.R. is grateful for financial support from the Austrian Science Fund (FWF) SFB ViCoM F4109-N13
and for computer support of the Vienna Scientific Cluster (VSC).
\end{acknowledgments}

\end{document}